\documentclass[conference]{IEEEtran}
\IEEEoverridecommandlockouts

\usepackage{cite}
\usepackage{adjustbox}
\usepackage{amsmath,amssymb,amsfonts}
\usepackage{algorithmic}
\usepackage{graphicx}
\usepackage{threeparttable}
\usepackage{textcomp}
\usepackage{xcolor}
\def\BibTeX{{\rm B\kern-.05em{\sc i\kern-.025em b}\kern-.08em
    T\kern-.1667em\lower.7ex\hbox{E}\kern-.125emX}}
    
\usepackage{lineno,hyperref}
\usepackage[ruled,vlined,english,linesnumbered]{algorithm2e}
\usepackage{breqn}
\usepackage{multirow}
\usepackage{capt-of}
\usepackage[T1]{fontenc}
\usepackage{letltxmacro}
\usepackage{thmtools}

\usepackage{rotating}
\modulolinenumbers[5]
\usepackage{multirow}
\usepackage{hyperref}
\usepackage{cleveref}
\usepackage{longtable}
\usepackage{caption}
\usepackage{subcaption}
\usepackage{makecell}

\usepackage{comment}    
\usepackage{todonotes}
\usepackage{tikz}
\newcommand*\circled[1]{\tikz[baseline=(char.base)]{
            \node[shape=circle,draw,inner sep=2pt] (char) {#1};}}
\newcommand{\aidoart}[1]{AIDOaRt}

\usepackage[normalem]{ulem}
\usepackage{todonotes}
\definecolor{todocolor}{rgb}{1.0,0.7,0.4}

\definecolor{notecolor}{rgb}{0.5,1.0,0.5}
\makeatletter
\define@key{todonotes}{note}[]{%
    \setkeys{todonotes}{color=notecolor}}%
\makeatother

\definecolor{darkocolor}{rgb}{0.1, 0.4, 0.8}

\definecolor{pierrecolor}{rgb}{0.8, 0.4, 0.1}

\definecolor{hamzehcolor}{rgb}{0.4, 0.1, 0.1}

\begin{document}

\title{AI-augmented Automation for Real Driving Prediction: an Industrial Use Case}


\author{\IEEEauthorblockN{Romina Eramo}
\IEEEauthorblockA{\textit{Dept. of Communication Science} \\
\textit{University of Teramo, Italy}\\
reramo@unite.it}
\and
\IEEEauthorblockN{Hamzeh Eyal Salman}
\IEEEauthorblockA{
\textit{Dept. of Software Engineering} \\
\textit{IT Faculty, Mutah University, Jordan}\\
hamzehmu@mutah.edu.jo}
\and
\IEEEauthorblockN{Matteo Spezialetti}
\IEEEauthorblockA{\textit{Dept. of Inf. Eng., Comp. Science and Math.} \\
\textit{University of L'Aquila, Italy}\\
matteo.spezialetti@univaq.it}

\and
\IEEEauthorblockN{Darko Stern}
\IEEEauthorblockA{\textit{Dept. of Research Program Management} \\
\textit{AVL, Austria}\\
darko.stern@avl.com}
\and
\IEEEauthorblockN{Pierre Quinton}
\IEEEauthorblockA{\textit{Dept. of Methodology and R\&D} \\
\textit{AVL, Austria}\\
pierre.quinton@avl.com}
\and
\IEEEauthorblockN{Antonio Cicchetti}
\IEEEauthorblockA{\textit{Dept. of IDT} \\
\textit{Mälardalen University, Sweden}\\
antonio.cicchetti@mdu.se}
}

\maketitle

\begin{abstract}
The risen complexity of automotive systems requires new development strategies and methods to master the upcoming challenges. Traditional methods need thus to be changed by an increased level of automation, and a faster continuous improvement cycle. In this context, current vehicle performance tests represent a very time-consuming and expensive task due to the need to perform the tests in real driving conditions. As a consequence, agile/iterative processes like DevOps are largely hindered by the necessity of triggering frequent tests.   

This paper reports on a practical experience of developing an AI-augmented solution - based on Machine Learning and Model-based Engineering
- to support continuous vehicle development and testing. In particular, historical data collected in real driving conditions is leveraged to synthesize a high-fidelity driving simulator and hence enable performance tests in virtual environments. Based on this practical experience, this paper also proposes a conceptual framework to support predictions based on real driving behavior. 
\end{abstract}

\begin{IEEEkeywords}
Continuous software engineering, DevOps, Machine Learning, Automotive, Real Driving Emission testing.
\end{IEEEkeywords}


\section{Introduction}\label{sec:intro}

Software plays an increasingly important role in modern vehicles. With the introduction of new features, new traffic regulations, system and security updates, automotive software requires the continuous verification and validation of new software versions, even after production. Thus, the traditional software life cycle, with slow feedback and manual interaction, is being replaced with a faster automatic feedback cycle and rapid continuous improvements. 

With the advent of DevOps principles \cite{Ebert2016,jabbari2016devops}, the (automotive) system engineering would benefit from supporting a continuous development involving a smooth continuum of information from design to runtime, and vice versa. Moreover, many leading companies have started to apply Artificial Intelligence (AI) principles and techniques for IT operations (AIOps)~\cite{aiops1, gartner-aiops}, to rethink the DevOps pipeline through continuous monitoring, alerting, and remediation securely and reliably. 
%
%

Models have a central role in vehicle development. Physical simulation models have been developed and optimized during the last decades and have reasonable maturity ~\cite{MDE}. Moreover, 
software-driven vehicle components are subject to constant further development, for instance, novel models to meet the forthcoming emission standards are needed. As a consequence, new methods to support full continuous software and system engineering processes are required. 
The ongoing European \emph{AIDOaRt} project\footnote{AIDOaRt ECSEL-JU project: \url{https://www.aidoart.eu/}} notably intends to address such issues. The project aims at providing a model-based framework to more efficiently support the continuous development of modern systems via AI-augmentation \cite{EramoMBBGBSC21, BRUNELIERE2022104672}.

This paper reports on a practical experience of developing an AI-augmented solution - based on Machine Learning (ML) -  to support continuous vehicle development and testing. The experience is based on a real industrial use case headed by AVL\footnote{AVL List GmbH, Graz (Austria) \url{http://www.avl.com/}}, one of the industrial partners of the AIDOaRt project. In particular, new challenges have been raised with the introduction of more stringent emissions legislation. In fact, the \emph{Real Driving Emissions} (RDE) test procedures \cite{RDE-EC} have been introduced in the EU aiming to evaluate nitrogen oxides (NOx) and particulate number (PN) emissions from passenger cars during on-road operation. In addition, more recent RDE legislation (euro7 compliance) extends the existing RDE criteria towards a much wider scope. In this context, AVL aims at improving the significance of test results including their evaluation in different vehicle development stages as well as the accuracy of simulation models. In fact, RDE test procedures tend to be time-consuming and expensive in real environments, and considering the needs for continuous development and testing they represent a bottle-neck in the process.  



In this paper, we present a conceptual approach that, starting from the modeling of the driver's behaviour, defines the core components of the solution for the prediction of RDE in virtual environments. The solution is based on model-based engineering (MBE) and ML: models are used to represent the necessary concepts that are included in the prediction; ML is used to generate high-fidelity simulations of the driver behaviour, and hence to enable the evaluation of driving emissions.

\smallskip

This paper is structured as follows. In section~\ref{sec:background}, we provide the necessary background to understand our proposal. Section~\ref{sec:approach} presents the real driving prediction framework. An application of this framework is provided in section~\ref{sec:ML}. Experimental results are discussed and evaluated in section~\ref{sec:evaluation}. Finally, the paper is concluded in section~\ref{sec:conclusion}.

\section{Background}\label{sec:background}
This section describes the basic concepts and context of the scope of the developed framework.

\subsection{Basic concepts}

\emph{Model Driven Engineering (MDE):} 
MDE allows raising the level of abstraction and thus improving the ability to engineer and handle complex systems~\cite{Thompson2018}.
The use of models as purposeful abstractions of systems and environments is also increasing within the industry (e.g., digital twining~\cite{Bordeleau2020}).
While first-generation MDE tools mainly focused on generating code from high-level models, they now also address model-based testing, verification, measurement, tool/language interoperability, or software evolution, among many other software engineering challenges. In system and software engineering, MDE contributes by 1) providing better abstraction principles and techniques (e.g., for the handled data), 2) facilitating the automation of engineering activities, and 3) supporting technology integration among all the covered design and development activities.

\emph{Artificial Intelligence and Machine Learning (AI/ML):}
The dissemination of Artificial Intelligence (AI), including Machine Learning (ML), principles and techniques in a regulated industry enables systems to decide and act in a more and more automated manner: it is used by companies to exploit the information they collect to improve the products and/or services they offer \cite{gartner-futureOfAI}. Lately, AI/ML is also impacting all aspects of the system and software development lifecycle, from specification to design, testing, deployment, and maintenance, with the main goal of helping engineers produce systems and software faster and with better quality while being able to handle ever more complex systems and software \cite{8812912, BurguenoKWZ21, felderer2023artificial}.

\subsection{The AIDOaRt approach}

\begin{figure}[t!]
	\centerline{\includegraphics[width=0.9\linewidth]{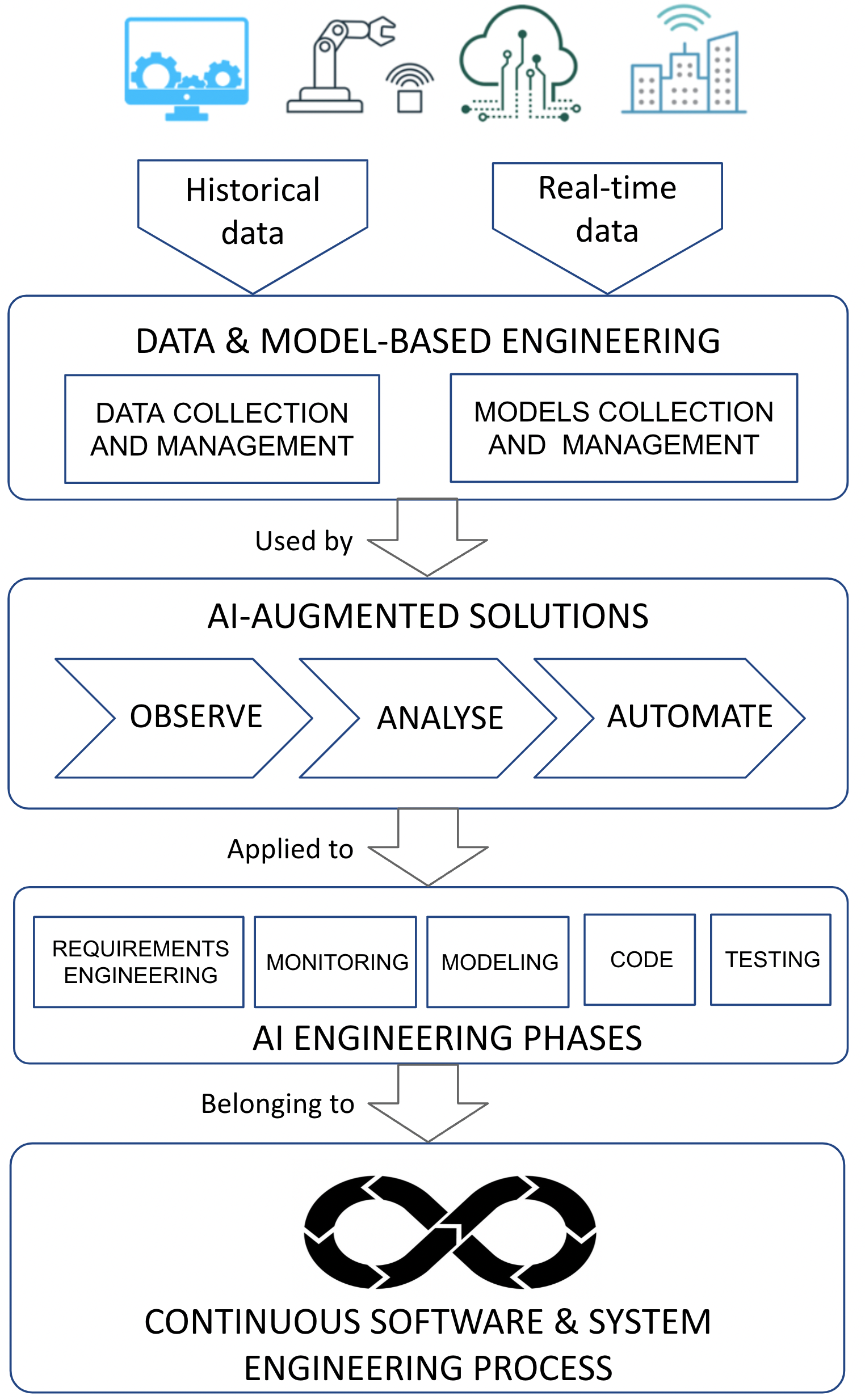}}
	\caption{\aidoart{} approach}
	\label{fig:concepts}
\end{figure}

Fig.~\ref{fig:concepts} provides a conceptual overview of the global solution based on the \aidoart{} project: it highlights the key principles and concepts that should be considered as the foundation of this work.

The overall component consumes different kinds of data, including runtime data (e.g., IT monitoring, log events, etc.) and design data produced during the software development process (e.g., software models, design documentation, traceability information, source code, etc.). All the collected data and models will be processed and stored; the \textit{Data \& Model-based Engineering} component is intended to support the standard DevOps practices by providing methods and tools for the data and models collection and management. The \textit{AI-augmented Solutions} component aims to enhance DevOps tool-chains (cf. existing DevOps tools~\cite{ceresani}) by employing AI and ML techniques in multiple \textit{Engineering Phases} of the system development process (e.g., requirements, monitoring, modeling, coding, testing, etc.). In an AIOps-enabled context, AI-augmented tools should support: 1) the monitoring of runtime data (such as logs, events, and metrics~\cite{2021_Valente}) and software data and their traceability (namely \textit{Observe}); 2) the analysis of both historical and real-time data (namely \textit{Analyze}); and 3) the automation of development and operation activities (namely \textit{Automate}). These capabilities will consume available design-time and runtime data that, according to MBE principles, should be made available to stakeholders as design-time and runtime models, respectively. 

The aim is to extend existing techniques and introduce novel solutions, enhancing the state of the art of, for instance, requirements engineering, monitoring, and testing, that already includes mechanisms supporting/leveraging data analysis~\cite{MNJR16, WMWH19, Briand2008}. Moreover, search-based techniques have been investigated to automate MBE-related activities such as language engineering, model transformation, and model versioning~\cite{BSAN17}. Nonetheless, especially when dealing with mission-critical systems, the automated generation of artifacts raises verification and validation issues, e.g., for certification purposes~\cite{LNPT18}. 

After the data acquisition and management/preprocessing, the AI-augmented solutions may provide different support to the DevOps pipelines. In this work, we propose a model-based and AI-augmented solution for predicting human driver behavior (addressing the component \emph{analysis} of the Figure) in the AVL RDE use case, described in detail in the next section. 

\subsection{The Real Driving Emissions case}\label{sec:rde}

\begin{figure}[hbt]
	\centering
	\includegraphics[width=1\linewidth]{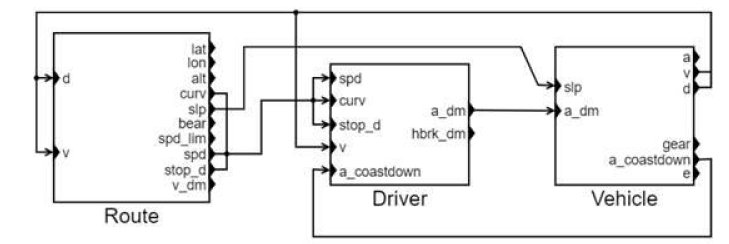} 
	\caption{AVL Route Studio Simulator architecture. \label{fig2:sim}}
\end{figure}

AVL is a large independent company that deals with the development, simulation, and testing of power-train systems and their integration into the vehicle. New challenges have been raised with the introduction of more stringent emissions legislation. {Particularly, procedures \cite{RDE-EC} for RDE testing have been introduced in the EU to evaluate nitrogen oxide (NOx) and particulate number (PN) emissions from passenger cars}. 
Comparing the vehicle performance against the prescribed regulations in real driving conditions is very time-consuming, subject to unforeseen testing conditions, and very expensive. Therefore, AVL is developing a high-fidelity driving simulator to reproduce the RDE test procedure in a virtual environment. The current AVL's Smart Mobile Solution - Route Studio\footnote{\url{https://www.avl.com/documents/10138/6781105/SMS_Simulation+Package_Solution+Sheet.pdf}} is composed of a set of models for the route, human driver, and vehicle (see Figure~\ref{fig2:sim}). The most critical component of the Route Studio Simulator is the human driver model since it needs to reproduce the behavior of a human driver as accurately as possible on the selected route for testing and with arbitrarily generated traffic conditions. 


\begin{figure}[]
	\centering
	\includegraphics[width=1\linewidth]{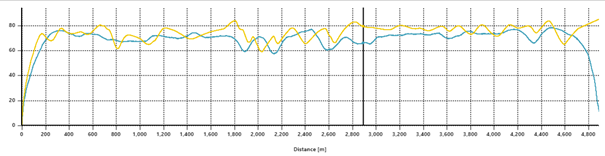} 
	\caption{Example of speed profile for a selected route (real driving in blue, simulated speed in yellow) \label{fig1:speedProfile}}
\end{figure}

In the current implementation, the behavior of the human driver is based on a rule-based parametric model, that due to modeling simplification lacks the fidelity required for high-fidelity RDE estimations. The deviations between a human driver and a rule-based parametric model mostly come from the fact that human drivers have more complex behaviors that are hard to encapsulate with a heuristic model.
Figure~\ref{fig1:speedProfile} show an example of a speed profile for both real driving (in blue) and a simulated one (in yellow) for a selected route.

In the AIDOaRt project collaboration, we aim to improve the fidelity of the human driver behaviour model by developing a data-driven model which simulates human-like driving on any arbitrary test route and under various traffic conditions. 
Such ML models will lead to a better estimate of engine exhaust emissions, and thereby reduce the burden and cost of assessing vehicle compliance with strict emissions legislation. 

%
%

\section{Real Driving Prediction framework}\label{sec:approach}


Figure~\ref{fig:fwk} presents a conceptual framework described using the MODA framework~\cite{9094197}, a conceptual modeling~\cite{MDE} framework that aims at supporting the description of data-centric systems in terms of models, data, and transformations. By proposing a conceptual framework, we do not discuss particular tools or technologies, but we categorize the different roles and the relationships of artefacts on a conceptual level. 
In particular, the figure shows a simplified class diagram for real driving prediction; it aims at designing components, classes, and their relationships for the prediction of driver behaviour. The objective of this framework is to enable predictions through the analysis of historical data and updated data.
It is composed of two main parts: i) the spatial model and predictors configuration (part I of Fig.~\ref{fig:fwk}) and ii) the domain-specific classes and relations describing the driving behaviour model we considered to use the framework (part II of Fig.~\ref{fig:fwk}) \cite{hal-03355162}.

\subsection{Model}

This package includes one or more models of the system under analysis or its parts (represented by the class \emph{Model}). In particular, it contains models that reflect the system and its environment in a descriptive manner, representing current or past aspects of the actual system, facilitating understanding, and enabling analysis \cite{EramoBCBWW22}.

\subsection{Prediction}

This package contains the abstract class \emph{Predictor} that encapsulates a (predictive) model used to predict information that has not been measured, allowing decision-making and trade-off analyses. This can include models for analysis, simulation, and ML (e.g., different types of ML algorithms like Random Forest, Ordinary Least Square, KNN, etc.). 

\subsection{Data}

This package contains different features and metrics definitions.
These are  \emph{Feature}, \emph{VolatileFeature}, and \emph{Metric}.

\subsection{Driving behaviour}

This package represents driving behaviours and related elements that determine such behaviours. The main elements are the environment, the vehicle, and the driver. The \emph{DrivingModel} class extends the class \emph{Model} and consists of at least one \emph{Driver}, one \emph{Vehicle}, and one \emph{Route}. In particular:

\begin{itemize}
    \item The settings for a specific driver are put in the \emph{Driver} class. Driver’s personal data such as \emph{reaction speed}, \emph{target velocity}, \emph{action}, and \emph{driving style}, will affect the driving behaviours. Driving style refers to \emph{normal driving}, \emph{zigzag driving}, \emph{risky acceleration}, or \emph{risky lane changing}. \emph{Reaction speed} and \emph{action attributes} refer to reactions toward other vehicles on the route. The \emph{action} attribute takes the following attributes: \emph{accelerating}, \emph{decelerating}, and \emph{maintaining} the current speed.
    \item To describe vehicles in detail, we introduce the \emph{Vehicle} class. It contains attributes that influence driving behaviours, including \emph{vehicle size}, \emph{weight, engine type, acceleration, and average emissions}. Note that the \emph{engineType} attribute is an enumeration literal, e.g. petrol, or hybrid.

   \begin{figure*}[]
	\centerline{\includegraphics[width=0.9\linewidth]{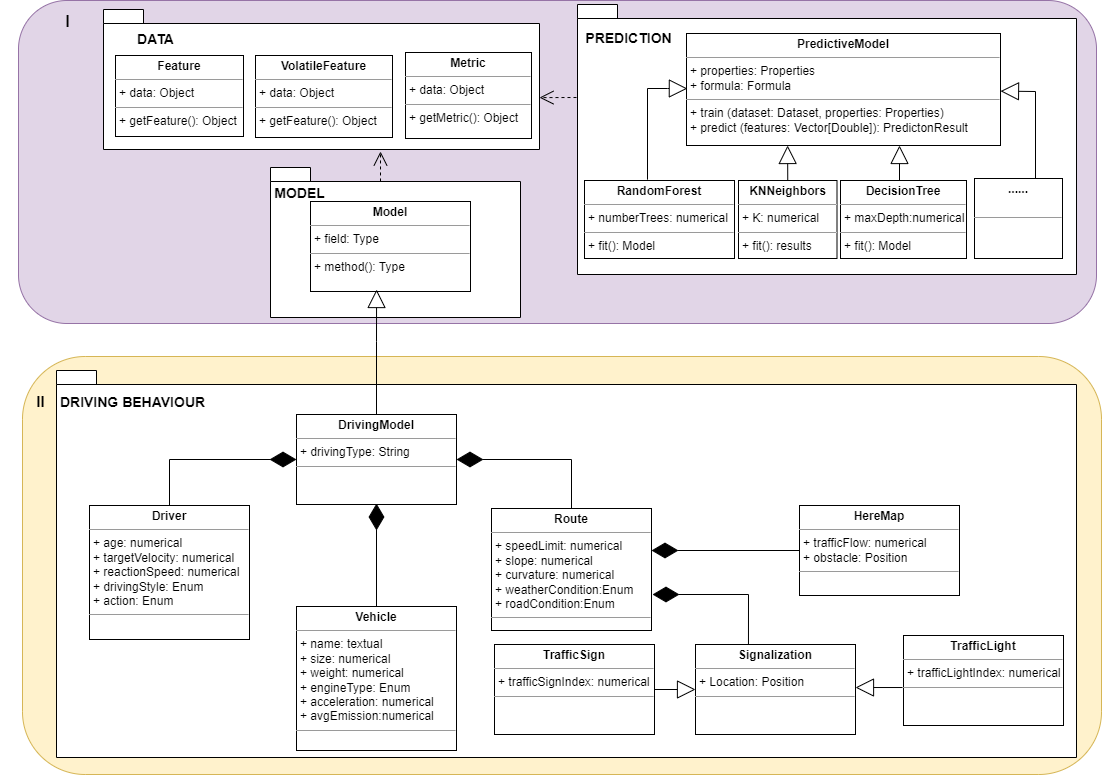}}
	\caption{Conceptual Framework for Real Driving Prediction}
	\label{fig:fwk}
\end{figure*}
    
    \item The \emph{Route} class is the entry point for configuring the environment for the driving scenario. It includes settings such as the \emph{speed limit}, \emph{slope}, \emph{curvature}, \emph{weather conditions} and \emph{road conditions}. The weather condition attribute takes the following values: \emph{slipperiness} and \emph{wind} while the road condition attribute has the following values: \emph{highway, urban, mountain}.
    Each route is composed of an \emph{HereMap} instance that allows customers to access a database of locations and map information. 
    Moreover, a route is composed of signalization details that can be represented by \emph{traffic signs} and \emph{traffic lights}, 
    \emph{road shapes}. 
\end{itemize}

This package represents a simplified version of a complete metamodel for defining driver behavior. However, it aims to be generic enough to suit many applications. In the next section, an application example is shown.
\section{Application to the RDE case}\label{sec:ML}

In this section, we present the application of the proposed framework on the RDE case, described in Sect.~\ref{sec:rde}. In particular, according to the RDE case providers, we focused on a simple case whose goal is to estimate the average behavior of drivers based on route data (see Route class in the framework). Thus, our goal is not to analyze individual driver behavior by considering, for instance, onboard sensors (see Driver and Vehicle classes in the framework) but rather to be able to estimate the average speed of drivers on a specific stretch of road, as the result affects emissions. 


\subsection{Data Acquisition}

Even if our experiment involved a subset of the dataset, we initially acquired all the data provided by AVL through AIDOaRT project
(see Table~\ref{tab:raw_features}). 
The data was collected from special equipment placed on a car. Data refers to several drivers and different driving paths: highways, mountains, etc. The data collected is real driving recordings in time series format (time-based data on \textit{vehicle speed, throttle/brake pedals, curvature, road gradient, GPS coordinates, etc.}). 
All of these data are continuous 1D signals with a sample rate of 1Hz up to 10Hz. 
The recorded data are then stored as a matrix of the form $M\times N$ where M is equal to the number of measured channels (features) and N is equal to the number of recorder values. 
 


\subsection{Features Selection}

To reduce the unnecessary computational cost, we filter out all the channels (features) that are not relevant for modeling human driver behavior. This filtering process is performed by the usecase provider based on its importance for driver behavior modeling.
To that end, we take a close look at the channels (features) in the dataset (see Table~\ref{tab:raw_features}) and found that there are three types of channels: \circled{1}  channels (resp. their corresponding data) are collected by two-way (
snapshots and real driving recordings),
  \circled{2} channels (volatile features like distance)  are calculated from other features (speed and time), and \circled{3} metric  channels (such as speed). 
The following channels are considered to train the data model while others are discarded: \emph{spd\_lim(16), tfc\_flw(17), traf\_lig(18), tfc\_sgn(19), toll\_booth(22), curvature(23), and slope(26)}. The \textit{velocity\_kmh\_raw(9)} feature is considered as a target (output). Numbers in parentheses after each channel refer to channel ID in Table~\ref{tab:raw_features}.

\begin{table}[]
\renewcommand{\arraystretch}{1.5}
\centering
\caption{Raw Real Driving Features}
\label{tab:raw_features}
\begin{threeparttable}[t]
\resizebox{\columnwidth}{!}{%
\begin{tabular}{|l|l|l|}
\hline
\textbf{ID} & \textbf{Feature}       & \textbf{Description}                                                            \\ \hline
1  & velocity\_raw & Raw imported vehicle speed                                           \\ \hline
2           & t                      & Time                                                                            \\ \hline
3           & speed\_raw             & Engine speed imported                                                           \\ \hline
4           & alt\_raw               & Altitude imported                                                               \\ \hline
5           & lat\_raw               & Latitude imported                                                               \\ \hline
6           & lon\_raw               & Longitude imported                                                              \\ \hline
7           & sat\_raw               & Nbr of satellite used for the measurement                                        \\ \hline
8           & d\_integrated\_raw     & Distance calculated out of the imported vehicle speed integration               \\ \hline
\textbf{9 \tnote{*}}           & \textbf{velocity\_kmh\_raw  }   & Raw imported vehicle speed                                                      \\ \hline
10          & d\_raw                 & Distance calculated out  of the latitude and the longitude                      \\ \hline
11          & lat                    & Snapped latitude                                                                \\ \hline
12          & lon                    & Snapped longitude                                                               \\ \hline
13          & alt                    & Altitude from Here Maps                                                         \\ \hline
14          & d                      & Distance out of  snapped longitude and latitude                                 \\ \hline
15          & here\_slope            & Slope from Here Maps                                                            \\ \hline
\textbf{16\tnote{*}} & \textbf{spd\_lim}      & \textbf{Speed limit from regulation}                                            \\ \hline
\textbf{17\tnote{*}} & \textbf{tfc\_flw}      & \textbf{Average speed from Here Maps}                                           \\ \hline
\textbf{18\tnote{*}} & \textbf{traf\_lig}     & \textbf{Traffic light index (until 6) to indicate the number of traffic lights} \\ \hline
\textbf{19\tnote{*}} & \textbf{tfc\_sgn}      & \textbf{Traffic sign index}                               \\ \hline
20          & sgn\_loc               & Localisation of the traffic sign (1=Left, 2=right, 3=above)                      \\ \hline
21          & conf                   & Confidence value from the snapping                                              \\ \hline
\textbf{22\tnote{*}} & \textbf{toll\_booth}   & \textbf{Index for toll booth}                                                   \\ \hline
\textbf{23\tnote{*}} & \textbf{curvature}     & \textbf{Road curvature in 1/m}                                                  \\ \hline
24          & curvature\_rad         & Road curvature in rad                                                           \\ \hline
25          & bearing                & Yaw of the vehicle                                                              \\ \hline
\textbf{26\tnote{*}} & \textbf{slope}         & \textbf{Slope calculated from the Here Maps altitude}           \\ \hline
27          & alt\_corr              & Corrected altitude\\ \hline
\end{tabular}
}
\begin{tablenotes}\footnotesize
\item[\textbf{*}] Selected features for training the prediction model
\end{tablenotes}
\end{threeparttable}
\end{table}

The discarded features are excluded for different reasons. Firstly, some features are derived or computed from other features. For example, distance can be computed using time and speed. Secondly, some features are repeated but are computed in different ways. For example, \emph{d, d_integrated_raw, d_raw,} represent the distance feature. Thirdly, some features have a negligible impact on the prediction, such as \emph{alt_raw, lat_raw, lon_raw, sat_raw}.

\subsection{Pre-processing}\label{sub:preprocessing}

When we explored the features' data in the dataset, we found that each feature's data has a different scale. For example,\textit{velocity\_kmh\_raw} feature ranges from $0$ to $144$ while values of \textit{tfc\_sgn} are integers ranging from $0$ to $42$. For this reason, we standardized each feature, except \textit{velocity\_kmh\_raw}, before proceeding with the training. 
The standard score ($Z_i$) of a value ($X_i$) is calculated as follows:

\begin{equation}\label{eq:standard}
Z_i = \frac{X_i-mean(X)}{Std\_Dev(X)}
\end{equation}
where  \textit{Mean}  and  \textit{Std\_Dev}  are  the  mean  and  the standard deviation of feature $X$, respectively.

The \textit{velocity\_kmh\_raw} feature represents the target feature or class. We decided to address the speed inference by posing the task as a classification problem. Therefore, since it ranges from $0$ to $144$ Km/h, we discretized it by dividing each data point $X_i$ by $10$ and rounding down the result to the closest integer. In this way, for example, any speed in the interval $[0,10)$ has been mapped into class $0$, any speed in the interval $[10,20)$ into class $1$, and so on.

\subsection{Driving Behaviour Prediction}

In the literature, there are many different ML algorithms for classification problems and there is no certain ML algorithm fit to all datasets. Choosing an ML algorithm depends on the type and size of the dataset of interest. 

In this research work, we compare the performance of mostly widely used ML algorithms using Python scikit-learn packages to determine which algorithm is the best fit for the collected dataset. These algorithms are \textit{GradientBoosting, DecisionTree, RandomForest, LogisticRegression, KNNeighbors, GaussianNB, LinearSVM, and AdaBoost}. The target algorithm should be able to predict the driver behaviour on the basis of the real dataset available and selected features on any arbitrary test route.

\section{Experimental Results and Evaluation}\label{sec:evaluation}

\subsection{Evaluation Procedure and Metrics}
The ML algorithms used in this research work are applied to the dataset, which is split into 90\% for training and 10\% for testing. For classification purposes, the splitting process is performed with a random shuffle, and we deal with the dataset not as time series data, accordingly. Also, the training data is split further into five portions (cv=5) in a process called cross-validation (CV for short). The following procedures are followed for each portion:

\begin{itemize}
\item [1-] ML model is trained using CV-1 portions.
\item [2-] The generated model is validated using the remaining portion.
\end{itemize}

The following standard evaluation metrics for classification solutions are used to evaluate the performance of these algorithms. The values of these metrics are scaled between 0 and 100\%. Our aim is to find an ML algorithm that maximizes these values. In these metrics, \textit{TP, TN, FP, FN} refers to \emph{true positive}, \emph{true negative}, \emph{false positive}, and \emph{false negative} prediction for each class, respectively.

\begin{equation}\label{eq:p}
Precision = \frac{TP}{TP+FP}
\end{equation}

\begin{equation}\label{eq:r}
Recall = \frac{TP}{TP+FN}
\end{equation}
\begin{equation}\label{eq:acc}
Accuracy = \frac{TP+TN}{TP+TN+FP+FN}
\end{equation}
\begin{equation}\label{eq:p}
F1-Score = \frac{2*Precision*Recall}{Precision+Recall}
\end{equation}

\begin{table*}

\renewcommand{\arraystretch}{1.5}
\centering
\caption{Cross validation results with CV=5 for all considered classifiers.}

\label{tab:CV_Results}
\begin{tabular}{|l|c|c|c|c|c|c|c|c|c|c|c|c|c|c|c|}
\hline
\textbf{}                       &
\multicolumn{2}{|c|}{\begin{sideways}  \textbf{GradientBoosting} \end{sideways}}&
\multicolumn{2}{|c|}{\begin{sideways}  \textbf{DecisionTree} \end{sideways}}&
\multicolumn{2}{|c|}{\begin{sideways}  \textbf{RandomForest} \end{sideways}}&
\multicolumn{2}{|c|}{\begin{sideways}  \textbf{LogisticRegression} \end{sideways}}&
\multicolumn{2}{|c|}{\begin{sideways}  \textbf{GaussianNB} \end{sideways}}&
\multicolumn{2}{|c|}{\begin{sideways}  \textbf{LinearSVM} \end{sideways}}&
\multicolumn{2}{|c|}{\begin{sideways}  \textbf{AdaBoost} \end{sideways}} \\ 
\hline
\hline
\textbf{} & AVG & STD & AVG &STD &AVG & STD&AVG   & STD &AVG  & STD & AVG& STD&AVG & STD \\ \hline
\textbf{Weighted Precision} &0.56  & 0.0 & 0.90 &0.0 &0.91 &0.0 & 0.28& 0.01  &0.19  &0.02 & 0.22&0.01 &0.17 &0.03 \\ \hline
\textbf{Weighted Recall} & 0.55 &  0.0& 0.90 &0.0 &0.91 &0.0 & 0.34&0.0 & 0.19   &0.03 &0.29 &0.0 & 0.19& 0.03\\ \hline
\textbf{Weighted F1-Score} & 0.54 &0.0  & 0.90 & 0.0&0.91 &0.0 & 0.26&0.0 & 0.10   &0.03 &0.19 &0.0 &0.13 & 0.02\\ \hline
\textbf{Weighted Accuracy} & 0.55 & 0.0 & 0.90 & 0.0& 0.91& 0.0& 0.34&0.0 &0.19   &0.03 &0.29 & 0.0& 0.19&0.03 \\ \hline

\end{tabular}
\end{table*}

\begin{table}[]
\renewcommand{\arraystretch}{1.7}
\centering
\caption{Obtained results using all considered classifiers.}
\label{tab:resultsComparison}
\resizebox{\columnwidth}{!}{%
\begin{tabular}{|l|c|c|c|c|c|c|c|c|}
\hline
\textbf{}                       &
\begin{sideways}  \textbf{GradientBoosting} \end{sideways}&
\begin{sideways}  \textbf{DecisionTree} \end{sideways}&
\begin{sideways}  \textbf{RandomForest} \end{sideways}&
\begin{sideways}  \textbf{LogisticRegression} \end{sideways}&
\begin{sideways}  \textbf{KNNeighbors}\end{sideways}&
\begin{sideways}  \textbf{GaussianNB} \end{sideways}&
\begin{sideways}  \textbf{LinearSVM} \end{sideways}&
\begin{sideways}  \textbf{AdaBoost} \end{sideways} \\ 
\hline
\hline
\textbf{Weighted AVG Precision} & \textbf{0.56}             & \textbf{0.91}         & \textbf{0.92}         & \textbf{0.25}               & \textbf{0.81}        & \textbf{0.18}       & \textbf{0.20}      & \textbf{0.18}     \\ \hline

\textbf{Weighted AVG Recall}    & \textbf{0.55}             & \textbf{0.91}         & \textbf{0.92}         & \textbf{0.34}               & \textbf{0.81}        & \textbf{0.15}       & \textbf{0.29}      & \textbf{0.20}     \\ \hline

\textbf{Weighted AVG F1-Score}  & \textbf{0.54}             & \textbf{0.91}         & \textbf{0.92}         & \textbf{0.26}               & \textbf{0.81}        & \textbf{0.07}       & \textbf{0.19}      & \textbf{0.15}     \\ \hline

\end{tabular}
}
\end{table}

We adjusted the hyperparameters of these algorithms to be the default values in the Python scikit-learn API. We used these default parameters without deliberate adjustment to ensure a fair comparison in the experiments among the ML algorithms. Also, these default values are often chosen by the library developers to work well in a wide range of scenarios.

\subsection{Results Discussion}

The dataset includes 27 features. Since some features have a minor impact on the target prediction, we selected the features listed in Table~\ref{tab:raw_features}. The selected data was used for the train prediction model of several popular ML algorithms (i.e., \textit{GradientBoosting, DecisionTree, RandomForest, LogisticRegression, KNNeighbors, GaussianNB, LinearSVM, and AdaBoost}). 


Table~\ref{tab:CV_Results} reports the performance results of the prediction models generated by the considered classifiers in the validation stage. As shown in this table, the prediction capability of \emph{RandomForest} and \emph{DecisionTree} classifiers is the best over other classifiers. The performance measures for these classifiers are [Precision: \%90 - \%91, Recall: \%90 - \%91, F1-score: \%90 - \%91, Accuracy: \%90 - \%91]. The KNN classifier is excluded from the validation stage. This is because there is no generated model by KNN classifier to validate. The KNN classifier depends on the majority rule for k nearest neighbours to classify unseen data~\cite{hassanat19}.

\begin{figure}[h!]
	\centering
	\includegraphics[width=1.0\linewidth]{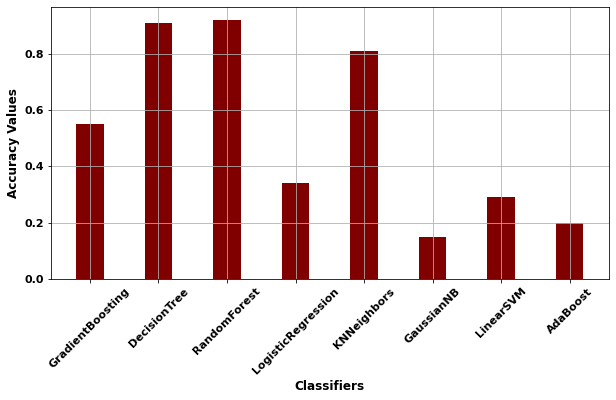} 
	\caption{Performance results of several classifiers.} \label{fig3:comparison}
\end{figure}

Table~\ref{tab:resultsComparison} lists performance results of prediction models built by ML algorithms of interest in terms of Precision, Recall, and F1-score. These algorithms were applied to entire dataset records. \textit{DecisionTree, RandomForest, and KNNeighbors} were the three top models among other classifiers where all metric values were greater than 90\% (for DecisionTree and RandomForest) and over 80\% for KNNeighbors. The highest metric values were produced by RandomForest as its values were 92\%. The worst metric values were produced by \textit{LogisticRegression, GaussianNB, LinearSVM, and AdaBoost} where their metric values were unable to achieve 50\%. The metric values of the remaining classifiers (\textit{GradientBoosting}) take a range between them. 

Figure~\ref{fig3:comparison} shows the comparison results among the classifiers of interest in terms of accuracy metric. Similar to what is noted in Table~\ref{tab:resultsComparison}, the figure shows the three top prediction models are built using \textit{DecisionTree, RandomForest, and KNNeighbors}. The highest accuracy results are produced by \textit{RandomForest} classifier. Also, the worst accuracy values were produced by \textit{LogisticRegression, GaussianNB, LinearSVM, and AdaBoost}.



Tables~\ref{tab:all_dataset_records}
shows detailed results of the RandomForest (RF) classifier, which is the best fit on the considered dataset. We worked on a multi-class classification problem; in fact, the dataset contains 15 classes, and each class represents a speed range (see preprocessing subsection~\ref{sub:preprocessing}). Although there is a large number of classes in the dataset, RF achieves high metrics values (Precision, Recall, F1-score, and Accuracy) for each class. RF provides consistent and encouraging results, as demonstrated by the accuracy values of 92\%. 


\begin{table}[]
\renewcommand{\arraystretch}{1.3}
\centering
\caption{Evaluation results of the \emph{RandomForest} classifier}
\label{tab:all_dataset_records}
\begin{tabular}{|ccccc|}
\hline
\multicolumn{1}{|c|}{\textbf{Class}}        & \multicolumn{1}{l|}{\textbf{Precision}} & \multicolumn{1}{l|}{\textbf{Recall}} & \multicolumn{1}{l|}{\textbf{F1-score}} & \multicolumn{1}{l|}{\textbf{Support}} \\ 
\hline
\hline
\multicolumn{1}{|c|}{0}                   & \multicolumn{1}{c|}{0.99}               & \multicolumn{1}{c|}{0.99}            & \multicolumn{1}{c|}{0.99}              & 8134                                  \\ \hline
\multicolumn{1}{|c|}{1}                   & \multicolumn{1}{c|}{0.94}               & \multicolumn{1}{c|}{0.93}            & \multicolumn{1}{c|}{0.93}              & 2498                                  \\ \hline
\multicolumn{1}{|c|}{2}                   & \multicolumn{1}{c|}{0.91}               & \multicolumn{1}{c|}{0.90}            & \multicolumn{1}{c|}{0.91}              & 4556                                  \\ \hline
\multicolumn{1}{|c|}{3}                   & \multicolumn{1}{c|}{0.90}               & \multicolumn{1}{c|}{0.90}            & \multicolumn{1}{c|}{0.90}              & 5984                                  \\ \hline
\multicolumn{1}{|c|}{4}                   & \multicolumn{1}{c|}{0.88}               & \multicolumn{1}{c|}{0.90}            & \multicolumn{1}{c|}{0.89}              & 7053                                  \\ \hline
\multicolumn{1}{|c|}{5}                   & \multicolumn{1}{c|}{0.88}               & \multicolumn{1}{c|}{0.87}            & \multicolumn{1}{c|}{0.88}              & 5339                                  \\ \hline
\multicolumn{1}{|c|}{6}                   & \multicolumn{1}{c|}{0.92}               & \multicolumn{1}{c|}{0.92}            & \multicolumn{1}{c|}{0.92}              & 3786                                  \\ \hline
\multicolumn{1}{|c|}{7}                   & \multicolumn{1}{c|}{0.93}               & \multicolumn{1}{c|}{0.92}            & \multicolumn{1}{c|}{0.92}              & 3072                                  \\ \hline
\multicolumn{1}{|c|}{8}                   & \multicolumn{1}{c|}{0.95}               & \multicolumn{1}{c|}{0.93}            & \multicolumn{1}{c|}{0.94}              & 2283                                  \\ \hline
\multicolumn{1}{|c|}{9}                   & \multicolumn{1}{c|}{0.93}               & \multicolumn{1}{c|}{0.93}            & \multicolumn{1}{c|}{0.93}              & 1511                                  \\ \hline
\multicolumn{1}{|c|}{10}                  & \multicolumn{1}{c|}{0.94}               & \multicolumn{1}{c|}{0.96}            & \multicolumn{1}{c|}{0.95}              & 1646                                  \\ \hline
\multicolumn{1}{|c|}{11}                  & \multicolumn{1}{c|}{0.84}               & \multicolumn{1}{c|}{0.81}            & \multicolumn{1}{c|}{0.83}              & 499                                   \\ \hline
\multicolumn{1}{|c|}{12}                  & \multicolumn{1}{c|}{0.78}               & \multicolumn{1}{c|}{0.79}            & \multicolumn{1}{c|}{0.79}              & 285                                   \\ \hline
\multicolumn{1}{|c|}{13}                  & \multicolumn{1}{c|}{0.88}               & \multicolumn{1}{c|}{0.82}            & \multicolumn{1}{c|}{0.85}              & 222                                   \\ \hline
\multicolumn{1}{|c|}{14}                  & \multicolumn{1}{c|}{0.64}               & \multicolumn{1}{c|}{0.47}            & \multicolumn{1}{c|}{0.54}              & 15                                    \\ \hline
\multicolumn{5}{|l|}{}                                                                                                                                                                                        \\ \hline
\multicolumn{1}{|l|}{\textbf{Accuracy}}     & \multicolumn{1}{c|}{\textbf{}}          & \multicolumn{1}{c|}{\textbf{}}       & \multicolumn{1}{c|}{\textbf{0.92}}     & \textbf{46883}                        \\ \hline
\multicolumn{1}{|c|}{\textbf{Weighted AVG}} & \multicolumn{1}{c|}{\textbf{0.92}}      & \multicolumn{1}{c|}{\textbf{0.92}}   & \multicolumn{1}{c|}{\textbf{0.92}}     & \textbf{46883}                        \\ \hline
\end{tabular}
\end{table}

\subsection{Observations and Lessons Learned}

We identified the following points as threats to the validity:

\begin{itemize}
    \item [-] Each driver's behavior differs from the other. For example, there are normal driving, aggressive driving, drowsy driving, etc. Our prediction model is built using only normal driving data as normal driving is the usual case.
    \item[-] In the current approach, we neglect the non-deterministic nature of environmental conditions like traffic conditions, due to the fact that it is not recorded when collecting data of the real driving cycle. In future work we plan to explore generative models such as variational autoencoders (VAE) or generative adversarial networks (GANs) to model environmental conditions.
\end{itemize}

\section{Related Work}\label{sec:related}
In this section, we present the most recent and relevant research works to this study. We categorize these works into three categories based on the type of algorithm used to predict the driver behaviour~\cite{8002632}: \textit{rule-based algorithms, ML-based algorithms, and digital twins.}

\subsection{Rule-based Driver Behavior prediction}
Rule-based algorithms are also called threshold-based algorithms. They are a set of algorithms that depend on the predefined threshold for monitored variables or factors to assign driver behavior to some class. In~\cite{Stoichkov13}, Radoslav proposed an approach based on thresholds computed using data collected from smart mobile sensors to detect the following deriving events: acceleration, deceleration, left turn, right turn, lane change to left, lane change to right. In~\cite{yii09}, Murphey et al. presented an approach based on the number of aggressive maneuvers to classify the driver behavior into: calm below 50\%,
aggressive above 100\%, and normal otherwise. Another classification for driver behavior is proposed by~\cite{andr13} and~\cite{mannt10}. This classification is based on fuel consumption or, in general, energy consumption. 

The main limitation of all previous-mentioned works is that they depend on a single parameter  and therefore the robustness and accuracy of the results are considerably limited. Another research direction relied on fuzzy logic to manage multiple parameters for driver behavior prediction. The research works in this direction are also based on predefined thresholds but are able to include
more parameters whilst keeping its simplicity, robustness, easy understanding and low computational order~\cite{dor14}\cite{GILM15}\cite{Syed27}\cite{jong03}. However, these works are also limited in terms of the number of
variables and data that can process. Also, they depend on thresholds computed using expert intervention.


\subsection{ML-based Driving Behavior Prediction}
A variety of ML-based driving behavior prediction approaches (with different purposes) are proposed. They employed different types of learning to achieve this prediction purpose, such as supervised learning, unsupervised learning, and combined unsupervised and supervised learning~\cite{8002632}\cite{ALIRAMEZANI2022100967}. In this section, we focus only on supervised learning approaches as they are the closest to our work presented in this article.  

The supervised learning approaches depend mainly on using driving data as training data to build a prediction model. In~\cite{s21196344}, Campoverde et al. proposed an approach to estimate emissions by applying ML to an important set of OBD data. The main aim of this approach is to determine the selected gear by the driver as the emission can be estimated based on the gear number used. This is achieved by integrating ANN, k-means clustering, and random forest algorithm. In~\cite{chen}, Chen et al. conducted an empirical study to evaluate the performance of different ML algorithms to classify the behavior of specific drivers, on the base of the sensor technology installed on the car. Although their purpose is different from ours (in fact, the authors want to recognize a specific driver based on the data collected), the results are interesting. They reported that there is no single machine-learning algorithm that fits all problems. Based on their obtained data, their experiments showed that the random forest approach is a good fit for identifying driving behaviors. In~\cite{Byttner},  Karginova et al. made a comparison between multiple ML algorithms for the purpose of classifying the driver's driving style. These algorithms are KNN, NN, decision tree, and random forest.  The KNN achieved the best performance within the nearest neighbor's inherent limits when clustering for K = 4 or 5.

\subsection{Digital Twins}
In~\cite{DYGALO2020121}, Dygalo et al. proposed to produce a digital twin of active vehicle safety systems for a proper simulation and correct system design. The digital twin consists of a set of modules ranked according to their priority. For example, “Wheel” and “Vehicle Body” modules were given top priority as they define the general parameters of the vehicle’s movement and location in physical space. This way of ranking modules within the system helps to detect inconsistencies in the top-priority modules at the earliest stages, thus tracking any errors and promptly correcting them.

\section{Conclusion and Future Work}\label{sec:conclusion}
This paper reported on a practical experience of developing an AI-augmented solution, based on Machine Learning and Model-based Engineering, to support continuous vehicle development and testing. We presented how historical data, collected in real driving conditions, is leveraged to synthesize a high-fidelity driving simulator and hence enable performance tests in virtual environments. Based on this practical experience, this paper also proposed a conceptual framework to support predictions based on real driving behavior.

In future work, we aim to extend the Driving Behaviour package
employing a complete metamodel describing the domain.
Moreover, we aim to propose a Digital Twin framework \cite{EramoBCBWW22}; in automotive, 
a digital twin of the product comprises the entire car, its software, mechanics, electrics, and physical behavior. This allows for simulation and validation of each step of the development to identify problems and possible failures before producing real parts.
We aim to focus on modeling the behavior of considered physical systems to make predictions by using several ML techniques. Also, we aim to validate the framework through several use cases in different domains (including the automotive and
RDE) by using different implementations.

\bibliographystyle{abbrv}
\bibliography{main}

\end{document}